\documentclass[11pt]{article}
\usepackage{hyperref}
\usepackage{cite}
\usepackage{footnote}
\makesavenoteenv{minipage}
\renewcommand{\title}[1]{%
    \bigskip%
    \begin{center}%
    \Large\bf #1%
    \end{center}%
    \vskip .2in}

\renewcommand{\author}[1]{%
    {\begin{center}
    #1
    \end{center}}}

\usepackage{amssymb,amsmath,amsfonts}
\usepackage{hyperref,cite}
\usepackage{color}

\newcommand{\EH}{Einstein-Hilbert }
\newcommand{\hg}{\hat{g}}
\newcommand{\pb}[1]{\left\{#1\right\}}
\newcommand{\cH}{\mathcal{H}}
\newcommand{\bcH}{\bar{\cH}}
\newcommand{\tcH}{\tilde{\cH}}
\newcommand{\cB}{\mathcal{B}}
\newcommand{\cD}{\mathcal{D}}
\newcommand{\cG}{\mathcal{G}}
\newcommand{\cM}{\mathcal{M}}
\newcommand{\cN}{\mathcal{N}}
\newcommand{\sR}{{}^{(3)}\!R}

\numberwithin{equation}{section}
\newcommand{\email}[1]{\footnote{E-mail: \href{mailto:#1}{#1}}}

\begin{document}
\title{BRST Quantization of Unimodular Gravity}

\author{Sudhaker Upadhyay$^{a}$\email{sudhakerupadhyay@gmail.com}~,
Markku Oksanen$^{b}$\email{markku.oksanen@helsinki.fi}~ and
Rodrigo Bufalo$^{c}$\email{rbufalo@ift.unesp.br}   \\
\textit{$^{a}$ \small Centre for Theoretical Studies, 
Indian Institute of Technology Kharagpur }\\
\textit{ \small  Kharagpur-721302,   India}\\
\textit{$^{b}$ \small Department of Physics, University of Helsinki, P.O. Box 64}\\
\textit{ \small FI-00014 Helsinki, Finland}\\
\textit{$^{c}$ \small Instituto de F\'{i}sica Te\'orica, Universidade Estadual Paulista} \\
\textit{\small Rua Dr. Bento Teobaldo Ferraz 271, Bloco II, 01140-070 S\~ao Paulo, SP, Brazil}\\
}

\begin{abstract}
We study the  quantization  of two versions of unimodular gravity,
namely, fully diffeomorphism-invariant unimodular gravity and unimodular
gravity with fixed metric determinant utilizing standard path integral approach.
We derive the BRST symmetry of effective actions corresponding to several relevant gauge
conditions. We observe that for some gauge conditions, the restricted
gauge structure may complicate the formulation and effective
actions, in particular, if the chosen gauge conditions involve the canonical momentum
conjugate to the induced metric on the spatial hypersurface. The BRST symmetry is extended
further to the finite field-dependent BRST transformation, in order to establish the
mapping between different gauge conditions in each of the two versions of
unimodular gravity.  
\end{abstract}
 \begin{flushleft}
\end{flushleft}


\section{Introduction}
\label{sec:intro}

Motivated by different purposes and scenarios a considerable attention
has been paid to alternative gravitational theories in recent years. In
particular, substantial efforts have been invested in understanding
the so-called cosmological constant problem
\cite{Weinberg:1988cp,Padmanabhan:2002ji,Bousso:2007gp}, more precisely
why the vacuum energy does not produce a huge value for the
cosmological constant, many orders of magnitude above the observed
value. Within this context a gravitational theory, nearly as old as
general relativity (GR) itself \cite{Einstein:1916vd}, the so-called
unimodular gravity (UG) \cite{Einstein:1919gv}, has once again been
analyzed \cite{Smolin:2009ti} as a potential way to approach the
problem.

Originally, the idea of unimodular gravity was conceived when Einstein
considered the unimodular condition \cite{Einstein:1916vd},
$\sqrt{-g}=1$, as a convenient way to partially fix a coordinate system
in GR.
The definition of unimodular gravity is usually based on the invariance
under a restricted group of diffeomorphisms that leave the determinant
of the metric invariant, so that the determinant of the metric can be
set equal to a fixed scalar density $\epsilon_0$, $ \sqrt{-g}=\epsilon_0
$.
Alternatively, one could consider restricted diffeomorphisms that
preserve the volume of spacetime \cite{Buchmuller:1988wx}.
The field equation for the metric is either the traceless Einstein
equation or, due to the Bianchi identity, the Einstein equation with a
cosmological constant \cite{Unruh:1988in}.

In comparison with GR, making the cosmological constant an arbitrary
constant of integration can be regarded as the key feature of unimodular
gravity. In order to achieve it, however, there is no need to constrain
the determinant of the metric. One can, therefore, either extend the
above unimodular condition in order to enlarge its group of symmetry,
e.g. by setting $\sqrt{-g}$ equal to the divergence of a vector density
field via parameterization of the spacetime coordinates
\cite{Kuchar:1991xd}. This kind of construction encompass the set of
theories known as fully diffeomorphism-invariant extensions of
unimodular gravity. The most prominent theories of this kind are the
Henneaux-Teitelboim theory \cite{Henneaux:1989zc} and the fully
diffeomorphism-invariant theory of unimodular gravity
\cite{Bufalo:2015wda}. The latter is no longer unimodular in the sense
that there is no condition on the determinant of the metric.
Nonetheless, it has been established how the theory is canonically
related to the conventional unimodular theory of gravity
\cite{Bufalo:2015wda}.

Returning to the aforementioned cosmological constant problem, a highly
speculative but interesting (formal) attempt to address this problem in
unimodular gravity has been made in \cite{Ng:1990rw,Smolin:2009ti} and
carefully revised in \cite{Bufalo:2015wda}, but with no decisive
conclusion. Unimodular gravity has also been used in investigating
other fundamental problems in gravitational theory. In particular, one
may argue that since the bulk part of the Hamiltonian of unimodular
gravity is nonvanishing, and the four-volume provides a cosmological
time, unimodular gravity could offer a new perspective on the problem of
time in quantum gravity and cosmology
\cite{Unruh:1988in,Sorkin:1987cd,Unruh:1989db}. However, later it was
shown that the problem of time persists in quantum unimodular gravity
\cite{Kuchar:1991xd}.

In classical level it is well known that unimodular gravity produces the
same physics as GR with a cosmological constant \cite{Unruh:1988in}.
However, a natural concern arises when such equivalence is investigated
in the quantum level, since a systematic and detailed study is necessary
and any conclusion beyond formal realm is always very subtle within
gravity. In addition, one may realize that quantization of each version
of unimodular gravity can be regarded as a potential quantization of GR.
Therefore, in order to shed a new light into several issues, analyses
considering the canonical structure and path integral quantization
\cite{Bufalo:2015wda} and radiative calculations
\cite{Alvarez:2015pla} of unimodular gravity have been
presented recently. Although very interesting and important conclusions
were drawn from such studies, several formal aspects still need to be
answered via deeper analysis within this scope. Hence, the
implementation of BRST formulations of the unimodular gravity theories
plays an interesting and important role in understanding the structure
of these theories. The BRST formulation is known to be a powerful method
for quantization of gauge theories, since it simplifies the proofs of
renormalizability, unitarity and anomaly cancellation.

A suitable approach for such analysis consist in an
extension of BRST symmetry realized by allowing the parameter to be
finite and
field-dependent, the so-called finite field-dependent BRST (FFBRST)
symmetry \cite{sdj}. Ref. \cite{sdj}  deals with the issue
 of generalizing BRST symmetry in Yang-Mills theories
 from the infinitesimal case to the finite case,
 while attempting to include the case of BRST-antiBRST
 symmetry by using the same approach as in the case of
 BRST symmetry, i.e., one that explicitly utilizes
 only a linear  dependence on the corresponding
 Grassmann-odd parameters.  The FFBRST symmetry transformations have
found several applications in a wide area of theoretical high energy
physics.

Within the most relevant results obtained from an analysis
following FFBRST symmetry we may cite, for instance, a correct
prescription for poles in the gauge field propagators in noncovariant
gauges has been derived with the help of FFBRST transformation by
connecting it to covariant gauges \cite{jog}. The long outstanding
problem of divergent energy integrals in Coulomb gauge has also been
regularized with the help of FFBRST transformation \cite{sdj1}.
The generalization of both on-shell and off-shell BRST as well as
anti-BRST symmetries for Yang-Mills theory are demonstrated explicitly
  where these are shown to establish the mapping between various gauges
of the theory \cite{susk}. The celebrated Gribov issue \cite{gri,
zwan,zwan1} has also been addressed by connecting the Yang-Mills theory
(possessing Gribov copies) to Gribov-Zwanziger theory (free from Gribov
copies within a Gribov
horizon) within the framework of FFBRST formulation (see refs. therein
\cite{sb}). The FFBRST transformations have been applied successfully in
the study of many other gauge theories \cite{smm,smm11,fs,sudd,fsm,
bl,g,abjm,rs}.

An extension of  FFBRST formulation has been established
for various theories at quantum level \cite{ssb,sud001} utilizing
Batalin-Vilkovisky (BV) formalism \cite{ht}. Lavrov and Lechtenfeld \cite{lav} suggests
 an alternative, w.r.t. \cite{sdj}, approach to generalize the
 BRST transformations in Yang-Mills theories,
 also by using a linear dependence on
 the corresponding Grassmann-odd parameter,
 naturally without having recourse to any
 quadratic dependence, since Ref. \cite{lav} does not
 deal with the case of BRST-antiBRST symmetry,
 and so any non-trivial quadratic dependence
 on the transformation parameters cannot occur.
  Moshin and Reshetnyak in Ref. \cite{ale00}  have
  systematically incorporated the case
 of BRST-antiBRST symmetry in Yang-Mills theories
 within the context of finite transformations
 that deals
 with the case of a quadratic  dependence
 on the corresponding parameters.
 This follows from the calculation
 of the corresponding Jacobian and from
 investigating the resulting quantum action.
The concept of finite BRST-antiBRST
 symmetry  is further extended    to the case of general
gauge theories \cite{ale1,mos1} as well as supersymmetric (SUSY) theories \cite{sual},
 whereas Ref. \cite{ale,mos}
 generalizes the corresponding parameters
 to the case of arbitrary   Grassmann-odd
 field-dependent parameters.
 We feel that the generalization of the BRST
formalism could be useful in understanding the quantization of
unimodular gravity theories.

The aim of the present paper is to investigate the features of  the two
unimodular gravity theories in the BRST as well as in generalized BRST
framework. Specifically, we discuss several potential gauge conditions
for the two unimodular gravity theories, one theory with
full diffeomorphism-invariance and the other with fixed metric
determinant. We compute the induced ghost action for each set of gauge
conditions, and write down the path integral for each effective action.
We demonstrate the nilpotent BRST symmetry of the effective action and
the corresponding transition amplitude. Moreover, we extend the BRST
symmetry by making the transformation parameter finite and field
dependent in the case of unimodular gravity. The action is invariant
under such a non-linear transformation of the fields. However, the
functional measure is not covariant under the FFBRST transformations. We
compute the non-trivial Jacobian for the functional measure under FFBRST
transformation for the two cases of unimodular gravity in general gauge
conditions. To illustrate this result we consider several gauge
conditions in both the fully diffeomorphism-invariant theory and the
theory with fixed metric determinant. Remarkably, we show that the
FFBRST transformation with certain parameters connects different gauges
of the given theories. In this way we are able to approach the different
sets of gauge conditions. Suppose any calculation in one set of gauge
conditions is unambiguous, a similar procedure for a different set of
gauge conditions could possibly be arrived at if one were to establish a
connection between the different sets of gauge conditions.

The paper is organized as follows. In section~\ref{sec1}, we discuss a
unimodular gravity theory extension endowed of fully
diffeomorphism-invariant theory in various gauge conditions. The
respective BRST symmetry transformations are derived and the gauge
fixing and ghost action is determined as well. A similar analysis for
unimodular gravity theory with a fixed metric determinant is presented
subsequently in section~\ref{sec2}. We analyse such theory in rather
different gauges than the full diffeomorphism invariant case. Further,
in section~\ref{sec3}, we provide a review of the methodology for the
FFBRST symmetry analysis in the case of fully diffeomorphism-invariant
unimodular gravity. We compute the explicit expression for Jacobian
under FFBRST transformation which depends on infinitesimal
field-dependent parameter explicitly. Under these circumstances, we show
that the Jacobian is responsible for the gauge connection between
different transition amplitudes. To be specific, we connect harmonic
gauge, synchronous gauge, axial gauge, Lorentz gauge and planar gauge to
each other for the fully diffeomorphism-invariant case. Nonetheless, the
unimodular Faddeev-Popov gauge, averaged metric determinant gauge and
averaged metric trace gauge are connected to each other in the fixed
metric determinant case. In the section~\ref{sec4} we summarize the
results.

 \section{Unimodular gravity with full diffeomorphism invariance }
\label{sec1}

We start our analysis with a brief review on the fully
diffeomorphism-invariant unimodular gravity. But first, it shows to be
convenient to revise the Henneaux-Teitelboim (HT) action
\cite{Henneaux:1989zc}
\begin{equation}
S_\mathrm{HT} =\int_{\cal M}d^4x \left(\frac{\sqrt{-g}R}{\kappa}
-\lambda(\sqrt{-g}-\partial_\mu \tau^\mu)
\right) +\oint_{\partial{\cal M}}d^3x
\left(\frac{2}{\kappa}\sqrt{|\gamma|}{\cal K}-\lambda r_\mu
{\tau^\mu} \right), \label{SHT}
\end{equation}
where $\tau^\mu$ is a \emph{vector density}, the gravitational coupling
constant is denoted as $\kappa=16\pi G$, $\gamma$ is the determinant of
the induced metric on the boundary $\partial\cM$ of spacetime, $\cal K$
is the extrinsic scalar curvature of $\partial\cM$, and $r_\mu$ is the
outward-pointing unit normal to the boundary $\partial\cM$. The
(fully diffeomorphism-invariant) unimodular condition has been
introduced into the action \eqref{SHT} as a constraint multiplied by a
Lagrange multiplier $\lambda$. The boundary term is included as in GR,
so that the variational principle for the action is well defined without
imposing boundary conditions on the derivatives of the metric.

The field equations consist of the Einstein equation, the equation for
the cosmological constant variable,
\begin{equation}
 \nabla_\mu\lambda=0,\label{nablalambda}
\end{equation}
a (fully diffeomorphism-invariant) unimodular condition,
\begin{equation}
 \sqrt{-g}=\partial_\mu\tau^\mu.\label{unimodcond.tau}
\end{equation}
The HT action \eqref{SHT} can indeed be derived from the UG action,
Eq.~\eqref{UG}, via parameterization of the spacetime coordinates
\cite{Kuchar:1991xd}.

We consider now an alternative action that is also fully
diffeomorphism-invariant and retains the classical equivalence with the
other unimodular theories. The action is written as
\begin{align}
 S_\mathrm{DUG}[g_{\mu\nu},\lambda,V^\mu ]&=\int_{\cM} d^4x\sqrt{-g}
 \left( \frac{R}{\kappa} -\lambda -V^\mu\nabla_\mu\lambda \right)
 +\frac{2}{\kappa}\oint_{\partial \cM}d^3x\sqrt{|\gamma|} {\cal K} ,
\label{SDUG}
\end{align}
where the variable $V^\mu$ is a \emph{vector field}. We shall refer to
this theory as the fully diffeomorphism-invariant unimodular gravity
(DUG). The action \eqref{SDUG} is arguably the most transparent
definition of  such a theory. The action \eqref{SDUG} consists of the
\EH action with a variable cosmological constant $\lambda$, and a
constraint term for
$\lambda$. The vector field $V^\mu$ acts as a Lagrange multiplier that
ensures $\nabla_\mu\lambda$ is zero in every direction, and thus
$\lambda$ is a constant. Classical solutions to the field equations
defined by the action \eqref{SDUG} are the same as for GR with a
cosmological constant.

The Hamiltonian analysis follows straightforwardly for the DUG action
when written in the Arnowitt-Deser-Misner (ADM) form
\cite{Bufalo:2015wda}. After a systematic canonical procedure at an
arbitrary gauge-fixing $\chi^{\mu}$, the path
integral for the given theory is found to be \cite{Bufalo:2015wda}
\begin{align}
 Z_\mathrm{DUG}={\cal N}^{-1}\int\prod_{x }\cD g_{\mu\nu}
g^{00}(-g)^{-\frac{3}{2}} N\delta(\chi^{\mu})
\left|\det\pb{\chi^{\mu},\cH_{\nu}}\right|  \exp\left(
\frac{i}{\hbar}S_\mathrm{EH}[g_{\mu\nu},\Lambda] \right),
 \label{ZDUG}
\end{align}
where we denoted the super-Hamiltonian and super-momentum constraints
collectively as
$\cH_{\nu}=(\cH_{T},\cH_{i})$ and $S_\mathrm{EH}$ is the \EH action with
a cosmological constant
\begin{align}
S_\mathrm{EH}[g_{\mu\nu}, \Lambda] =\frac{1}{\kappa} \int_{\cal M} d^4x
\sqrt{-g}(R-2\Lambda)
+\frac{2}{\kappa}\oint_{\partial {\cal M}} d^3x
\sqrt{\left|\gamma\right|}{\cal K}.
\end{align}
It should be noted that the value of $\Lambda$ is not set by the action.
The cosmological constant $\Lambda$ is an unspecified value of the
variable $\lambda$.

The present theory has the advantage of enabling the use of the same
(covariant) gauges for the diffeomorphism symmetry as in GR. In view of
this, and bearing in mind the BRST analysis, let us recall that the
infinitesimal (diffeomorphism) gauge transformation of the metric is
written as
\begin{equation}\label{infDiff}
 \delta_\xi g_{\mu\nu}=\partial_\rho g_{\mu\nu}\xi^\rho
 +g_{\mu\rho}\partial_\nu\xi^\rho+g_{\rho\nu}\partial_\mu\xi^\rho.
\end{equation}
The inverse metric density is defined as
\begin{equation}\label{hg}
 \hg^{\mu\nu}=\sqrt{-g}g^{\mu\nu},
\end{equation}
and its transformation under \eqref{infDiff} is obtained as
\begin{equation}
 \delta_\xi\hg^{\mu\nu}=\partial_\rho(\hg^{\mu\nu}\xi^\rho)
 -\hg^{\mu\rho}\partial_\rho \xi^\nu  -\hg^{\rho\nu}\partial_\rho
\xi^\mu.
\end{equation}

\subsection{BRST Symmetry}

The BRST transformation for the full set of fields, metric field
$g_{\mu\nu}$, Faddeev-Popov ghost fields $c^\mu, \bar{c}_\nu$, and
Nakanishi-Lautrup auxiliary field $\eta_\mu$, can be obtained from the
properties of infinitesimal diffeomorphisms as
\begin{subequations}
\begin{align}
 \delta_b g_{\mu\nu}&=\left( \partial_\rho g_{\mu\nu}c^\rho
 +g_{\mu\rho}\partial_\nu c^\rho+g_{\rho\nu}\partial_\mu c^\rho \right)
 \theta, \label{BRSTtransa}\\
 \delta_b c^\mu&=-c^\nu\partial_\nu c^\mu\theta, \label{BRSTtransb}\\
 \delta_b\bar{c}_\mu&=\eta_\mu\theta, \label{BRSTtransc}\\
 \delta_b\eta_\mu&=0.\label{BRSTtransd}
\end{align}
\end{subequations}
The inverse metric density \eqref{hg} transforms under
\eqref{BRSTtransa} as
\begin{equation}
 \delta_b\hg^{\mu\nu}=\left( \partial_\rho(\hg^{\mu\nu}c^\rho)
 -\hg^{\mu\rho}\partial_\rho c^\nu
 -\hg^{\rho\nu}\partial_\rho c^\mu\right)\theta.
\end{equation}
The BRST transformation of the metric is obtained from the infinitesimal
diffeomorphism \eqref{infDiff}, with the replacement
$\xi^\rho\rightarrow c^\rho\theta$. The transformation of the ghost
$c^\mu$ was obtained from the commutator of vector fields generating the
infinitesimal diffeomorphisms by replacing the vector components with an
anticommuting field: $(c=c^\mu\partial_\mu)$
\begin{equation}
 -\frac{1}{2}[c,c]^\mu=-\frac{1}{2}( c^\nu\partial_\nu c^\mu
 -\partial_\nu c^\mu c^\nu)=-c^\nu\partial_\nu c^\mu.
\end{equation}
The transformation of the anti-ghost $\bar{c}_\mu$ is proportional to
the auxiliary field $\eta_\mu$ that acts as a Lagrange multiplier of
gauge conditions. The transformations
\eqref{BRSTtransa}-\eqref{BRSTtransd} commute with spacetime
differentiation.

\subsection{Gauge fixing and ghost action }

Next we derive the BRST invariant gauge fixing and ghost action
$S^G_{gf+gh}$ for different sets of gauge conditions, determining thus
the respective path integral expression. Moreover, as aforementioned, we
shall restrict our discussion to covariant and one non-covariant gauges
for the DUG theory, while for the UG theory we will consider only
non-covariant gauges.

\subsubsection{Harmonic gauge}

Let us start our analysis by choosing the transverse harmonic gauge,
\begin{equation}\label{gauge.H}
 \partial_\nu\hg^{\mu\nu}=0.
\end{equation}
The gauge and ghost action can be written in the form
\begin{equation}\label{SH}
 S^H_{gf+gh}=\int d^4x\left( -\eta_\mu\partial_\nu\hg^{\mu\nu}
 +\partial_\nu\bar{c}_\mu \left(
\partial_\rho(\hg^{\mu\nu}c^\rho)
 -\hg^{\mu\rho}\partial_\rho c^\nu
 -\hg^{\rho\nu}\partial_\rho c^\mu \right) \right).
\end{equation}
In the action \eqref{SH}, the terms that involve the gauge conditions
\eqref{gauge.H} could be absorbed into the gauge-fixing terms via a
shift transformation of the auxiliary fields $\eta_\mu$. Still we choose
to keep those terms in order to maintain manifest BRST invariance. Thus,
we find the path integral in the harmonic gauge
\begin{align}
Z^H_\mathrm{DUG} ={\cal N}^{-1}\int\prod_x{\cal D}g_{\mu\nu}{\cal
D}\eta_\mu {\cal D}\bar{c}_\mu {\cal D}c^\nu g^{00}(-g)^{-\frac{3}{2}}
\exp\left(\frac{i}{\hbar}\left[ S_{EH}[g_{\mu\nu}, \Lambda]+
S^H_{gf+gh} \right]\right). \label{PIH}
\end{align}

\subsubsection{Lorentz covariant $\alpha$-gauge}

A direct generalization of the above condition is the Lorentz covariant
$\alpha$-gauge
\begin{equation}\label{gauge.L}
 \partial_\nu\hg^{\mu\nu}+\alpha\hg^{\mu\nu}_\mathrm{R}\eta_\nu=0,
\end{equation}
where $\hg^{\mu\nu}_R$ is a fixed reference background metric density.
The limit $\alpha\rightarrow0$ reproduces the harmonic gauge. The gauge
and ghost action with an arbitrary constant parameter $\alpha$ is
written as
\begin{equation}
 S^\alpha_{gf+gh}=\int d^4x\left(
 -\frac{\alpha}{2}\hg^{\mu\nu}_\mathrm{R}\eta_\mu\eta_\nu
 -\eta_\mu\partial_\nu\hg^{\mu\nu}
 +\partial_\nu \bar{c}_\mu \left(
 \partial_\rho(\hg^{\mu\nu}c^\rho)
 -\hg^{\mu\rho}\partial_\rho c^\nu
 -\hg^{\rho\nu}\partial_\rho c^\mu \right) \right),
\end{equation}
which is similar to the action obtained in GR \cite{Nishijima:1978wq}.
Hence, the BRST invariant path integral in the $\alpha$-gauge reads
\begin{equation}
Z^\alpha_\mathrm{DUG} ={\cal N}^{-1}\int\prod_x{\cal D}g_{\mu\nu}
{\cal D}\eta_\mu{\cal D}\bar{c}_\mu {\cal D}c^\nu
g^{00}(-g)^{-\frac{3}{2}}
\exp \left(\frac{i}{\hbar} \left[ S_{EH}[g_{\mu\nu}, \Lambda]+
S^\alpha_{gf+gh} \right] \right).\label{fd}
\end{equation}

\subsubsection{Axial gauge}

A well-known condition by computation purposes is the axial gauge. This
condition is suitable, in particular, due to the fact that the ghost
fields are decoupled and can simply be dropped. It reads
\begin{equation}\label{gauge.A}
 a_\nu\hg^{\mu\nu}=0,
\end{equation}
where $a_\nu$ is a fixed one-form. The gauge and ghost action can be
written in the following form
\begin{equation}
 S^A_{gf+gh}=\int d^4x\left( -a_{(\mu}\eta_{\nu)}\hg^{\mu\nu}
 -a_{(\mu}\bar{c}_{\nu)}\left[ \partial_\rho(\hg^{\mu\nu}c^\rho)
 -\hg^{\mu\rho}\partial_\rho c^\nu  -\hg^{\rho\nu}\partial_\rho c^\mu
\right]\right),
\end{equation}
and, finally, we find the path integral in the axial gauge as
\begin{equation}
Z^A_\mathrm{DUG} ={\cal N}^{-1}\int\prod_x{\cal D}g_{\mu\nu}
{\cal D}\eta_\mu{\cal D}\bar{c}_\mu {\cal D}c^\nu
g^{00}(-g)^{-\frac{3}{2}}
\exp \left(\frac{i}{\hbar}\left[ S_{EH}[g_{\mu\nu}, \Lambda]+
S^A_{gf+gh} \right]\right). \label{PIA}
\end{equation}

\subsubsection{Planar gauge}

Again, we can consider an extension, the planar gauge, by introducing to
the axial gauge an arbitrary constant parameter $\alpha$ such as
\begin{equation}\label{gauge.P}
 a_\nu\hg^{\mu\nu} +\alpha\hg^{\mu\nu}_\mathrm{R}\eta_\nu=0.
\end{equation}
The limit $\alpha\rightarrow0$ reproduces the axial gauge. The BRST
invariant gauge and ghost action is written in the form
\begin{equation}
 S^P_{gf+gh}=\int d^4x\left(
-\frac{\alpha}{2}\hg^{\mu\nu}_\mathrm{R}\eta_\mu\eta_\nu
-a_{(\mu}\eta_{\nu)}\hg^{\mu\nu}
 -a_{(\mu}\bar{c}_{\nu)}\left[ \partial_\rho(\hg^{\mu\nu}c^\rho)
 -\hg^{\mu\rho}\partial_\rho c^\nu  -\hg^{\rho\nu}\partial_\rho c^\mu
\right]\right).
\end{equation}
We thus find the following expression for the path integral in the
planar gauge
\begin{equation}
Z^P_\mathrm{DUG} ={\cal N}^{-1}\int\prod_x{\cal D}g_{\mu\nu}
{\cal D}\eta_\mu{\cal D}\bar{c}_\mu {\cal D}c^\nu
g^{00}(-g)^{-\frac{3}{2}}
\exp \left(\frac{i}{\hbar} \left[ S_{EH}[g_{\mu\nu}, \Lambda]+
S^P_{gf+gh} \right]\right). \label{genp}
\end{equation}

\subsubsection{Synchronous gauge}

By means of complementarity let us consider another well-known
condition, the synchronous gauge. It reads
\begin{equation}\label{gauge.S}
 \chi_0=g_{00}+1=0,\quad \chi_i=g_{0i}=0,
\end{equation}
where $i=1,2,3$.
We now obtain a non-covariant expression for the gauge and ghost action
\begin{align}
 S^S_{gf+gh}&=\int d^4x\sqrt{-g}\left[ -\eta^0(g_{00}+1)-\eta^ig_{0i}
 -\bar{c}^0\nabla_\mu c^\mu \right.\nonumber\\
&\quad -\left.\bar{c}^\mu\left( g_{0\mu}\nabla_\nu c^\nu  +\partial_\nu
g_{0\mu} c^\nu +g_{0\nu}\partial_\mu c^\nu  +g_{\mu\nu}\partial_0 c^\nu
\right) \right],
\end{align}
where
\begin{equation}
 \nabla_\mu c^\mu=\partial_\mu c^\mu
+\frac{1}{2}g^{\mu\nu}\partial_\rho g_{\mu\nu}c^\rho.
\end{equation}
Finally, the path integral in this gauge is written as
\begin{align}
Z^S_\mathrm{DUG} ={\cal N}^{-1}\int\prod_x{\cal D}g_{\mu\nu}
{\cal D}\eta_\mu{\cal D}\bar{c}_\mu {\cal D}c^\nu
g^{00}(-g)^{-\frac{3}{2}}
\exp \left(\frac{i}{\hbar}\left[ S_{EH}[g_{\mu\nu}, \Lambda]+
S^S_{gf+gh}\right] \right). \label{PIS}
\end{align}

With this last study we conclude the first analysis by discussing the
BRST invariant approach for the DUG theory. This allowed us to
determine consistently the respective gauge fixing and ghost action,
and subsequently the transition amplitude, for a series of gauge
conditions. We shall now extend this study to the UG theory.

\section{Unimodular gravity with fixed metric determinant}
\label{sec2}

Once the BRST analysis of the UG theory will resort to subtle points of
the Hamiltonian analysis \cite{Bufalo:2015wda}, we shall make a brief
review of relevant aspects of the Hamiltonian analysis of UG. The
standard approach to define UG is to introduce the unimodular condition
into \EH action as a constraint multiplied by a Lagrange multiplier
$\lambda$,
\begin{equation}\label{UG}
S_\mathrm{UG}= \int_{\cal M}d^4x
\left(\frac{\sqrt{-g}R}{\kappa}-\lambda(\sqrt{-g}-\epsilon_0)\right)
+\frac{2}{\kappa}\oint_{\partial{\cal M}}d^3x\sqrt{|\gamma |}{\cal K}.
\end{equation}
where $\epsilon_0$ is a fixed \emph{scalar density}, such that
$\epsilon_0d^4x$ defines a proper volume element. Then we introduce the
ADM variables. The above action is written in
ADM form as
\begin{equation}
S_\mathrm{UG}=\int dt\int_{\Sigma_t}
\left[\frac{N\sqrt{h}}{\kappa}(K_{ij}{\cal G}^{ijkl}K_{kl} +{}^{(3)}R)-
\lambda(N\sqrt{h}- \epsilon_0)  \right]+S_{\cal B},
\end{equation}
where $N$ is the lapse variable and $N^i$ is the shift vector on the
spacelike hypersurface $\Sigma_t$, the extrinsic curvature $ K_{ij}$ is
written as
\begin{equation}\label{Kij}
 K_{ij}=\frac{1}{2N}\left( \partial_th_{ij}-D_iN_j-D_jN_i \right),
\end{equation}
where $D$ is the covariant derivative that is compatible with the
(induced) metric $h_{ij}$ on $\Sigma_t$, and $h^{ij}$ is the inverse
metric, $h_{ij}h^{jk}=\delta_i^{k}$, and the boundary contribution
$S_{\cB}$ is given as in GR.

The Hamiltonian analysis leads to the following path integral in the
$\tilde\chi^{\mu}$ gauge condition \cite{Bufalo:2015wda},
\begin{align}\label{ZUG}
 Z_\mathrm{UG}&=\cN^{-1}\int\prod_{x^\mu}\cD g_{\mu\nu}
g^{00}(-g)^{-\frac{3}{2}}  \delta\left( \frac{\int_{\Sigma_t}\left(
\sqrt{-g}-\epsilon_0 \right)}
 {\int_{\Sigma_t}\sqrt{h}} \right) \nonumber \\
& \quad \times N\delta(\tilde\chi^{\mu})
\left|\det\pb{\tilde\chi^{\mu},\tcH_{\nu}}\right|
\exp\left(\frac{i}{\hbar}S_\mathrm{EH}[g_{\mu\nu}] \right).
\end{align}
It should be noted that the $\delta$-function imposes the unimodular
condition to hold on each slice $\Sigma_t$ of spacetime in average,
$\int_{\Sigma_t}(\sqrt{-g}-\epsilon_0)=0$.

In view of the BRST symmetry, let us recall some subtle points involving
the gauge generators of UG. In unimodular gravity with fixed metric
determinant, the ADM gauge transformation of a function $\varphi$ of the
canonical variables $h_{ij}$ and $\pi^{ij}$ is given as
\begin{equation}\label{gt.uADM}
 \delta_{\tilde\xi}\varphi=\pb{\varphi,\int_{\Sigma_t}
 \tcH_\mu\tilde\xi^\mu},  \quad
\tcH_\mu\tilde\xi^\mu=\bcH_T\bar\xi+\cH_i\xi^i,
\end{equation}
where the gauge parameter $\tilde\xi^\mu$ consists of an average-free
scalar and a
three-vector, $\tilde\xi^\mu=(\bar\xi,\xi^i)$,
$\int_{\Sigma_t}\sqrt{h}\bar\xi=0$, and the generators are
the first class (average-free) Hamiltonian and super-momentum
constraints $\tcH_\mu=(\bcH_T,\cH_i)$,
\begin{align}\label{bcH_T}
 \bcH_T&=\overline{\frac{\kappa}{\sqrt{h}}\pi^{ij}\cG_{ijkl}\pi^{kl}}
 -\overline{\frac{\sqrt{h}}{\kappa}\sR}\approx0,\\
 \cH_i&=-2h_{ij}D_k\pi^{jk}\approx0,
\end{align}
where the overline denotes average-free components, whose integral over
space vanish, defined as
\begin{align}
 \overline{\frac{\kappa}{\sqrt{h}}\pi^{ij}\cG_{ijkl}\pi^{kl}}&=
 \frac{\kappa}{\sqrt{h}}\pi^{ij}\cG_{ijkl}\pi^{kl}
 -\frac{\sqrt{h}}{\int_{\Sigma_t}\sqrt{h}} \int_{\Sigma_t}
 \frac{\kappa}{\sqrt{h}}\pi^{ij}\cG_{ijkl}\pi^{kl}, \\
 \overline{\sqrt{h}\sR}&=\sqrt{h}\sR
 -\frac{\sqrt{h}}{\int_{\Sigma_t}\sqrt{h}} \int_{\Sigma_t}\sqrt{h}\sR.
\end{align}
Since the zero mode of the Hamiltonian constraint
\begin{equation}\label{cH0}
 \cH_0=\int_{\Sigma_t}\cH_T=\int_{\Sigma_t}\left(  \frac{\kappa}
{\sqrt{h}}\pi^{ij}\cG_{ijkl}\pi^{kl}  -\frac{\sqrt{h}}{\kappa}\sR
\right)  +\lambda_0\int_{\Sigma_t}\sqrt{h}\approx0
\end{equation}
is a second class constraint in the present theory, it does not generate
a gauge transformation.

The average-free gauge parameter $\bar\xi$ depends of the metric so that
it remains average-free under a variation of the metric,
\begin{equation}
 \delta\int_{\Sigma_t}\sqrt{h}\bar\xi=\int_{\Sigma_t}\left(
 \delta\sqrt{h}\bar\xi+\sqrt{h}\delta\bar\xi \right)=0.
\end{equation}
This implies that the gauge parameter $\bar\xi$ can be expressed as
\begin{equation}
 \bar\xi=\xi-\xi_0,\quad \xi_0=\frac{1}{\int_{\Sigma_t}\sqrt{h}}
 \int_{\Sigma_t}\sqrt{h}\xi,
\end{equation}
where $\xi$ is an unrestricted field that does not depend on any
variable. Now the identity $\int_{\Sigma_t}\sqrt{h}\bar\xi=0$ can be
used anywhere, even inside of Poisson brackets. On the other hand, it
means that $\bar\xi$ has a nonvanishing Poisson bracket with the
canonical momentum $\pi^{ij}$.

In the ADM gauge transformation \eqref{gt.uADM} we can write the
average-free part of the generator as
\begin{equation}
 \int_{\Sigma_t}\bcH_T\bar\xi
=\int_{\Sigma_t}\cH_T^\mathrm{GR}\xi
-\cH_0^\mathrm{GR}\xi_0,
\end{equation}
where $\cH_0^\mathrm{GR}=\int_{\Sigma_t}\cH_T^\mathrm{GR}$ and
\begin{align}
 \cH_T^\mathrm{GR}&=\frac{\kappa}{\sqrt{h}}\pi^{ij}\cG_{ijkl}\pi^{kl}
 -\frac{\sqrt{h}}{\kappa}\sR. \label{cHTGR}
\end{align}
Note that $\cH_T^\mathrm{GR}$ and $\cH_0^\mathrm{GR}$ are not
constraints in the present theory, since they do not include the
cosmological term $\sqrt{h}\lambda_0$ (see \eqref{cH0}). Actually, we
shall use the following equivalent form of the full generator of
the gauge transformations \eqref{gt.uADM}
\begin{equation}\label{ggen2.uADM}
 \int_{\Sigma_t}\tcH_\mu\tilde\xi^\mu=  \int_{\Sigma_t}\left(
\cH_T^\mathrm{GR}\bar\xi+\cH_i\xi^i \right),
\end{equation}
since it avoids the appearance of $\cH_0^\mathrm{GR}$ in evaluation of
the transformations.

Gauge transformation of canonical variables are obtained from
\eqref{gt.uADM} as follows. The spatial metric transforms as
\begin{equation}\label{gt.uADM.metric}
 \delta_{\tilde\xi}h_{ij}=\frac{2\kappa}{\sqrt{h}}\cG_{ijkl}\pi^{kl}
 \bar\xi  +\partial_kh_{ij}\xi^k +h_{ik}\partial_j\xi^k
+h_{kj}\partial_i\xi^k,
\end{equation}
since $\pb{h_{ij},\bar\xi}=0$. The canonical momentum $\pi^{ij}$
transforms as
\begin{align}\label{gt.uADM.pi}
 \delta_{\tilde\xi}\pi^{ij}&=\bigg[ \frac{1}{2}h^{ij}\left(
\frac{\kappa}{\sqrt{h}}\pi^{kl}\cG_{klmn}\pi^{mn}
+\frac{\sqrt{h}}{\kappa}\sR \right)
 -\frac{\kappa}{\sqrt{h}}\left( 2\pi^{(i}_{\ \:k}\pi^{j)k}
 -\pi^{ij}h_{kl}\pi^{kl} \right)  \nonumber \\
&\quad - \frac{\sqrt{h}}{\kappa}\left( \sR^{ij} -D^iD^j +h^{ij}D^kD_k
  \right) \bigg] \bar\xi +\partial_k\left( \pi^{ij}\xi^k \right)
-\pi^{ik}\partial_k\xi^j  -\pi^{kj}\partial_k\xi^i \nonumber \\
 &\quad +\left( \frac{1}{\int_{\Sigma_t}\sqrt{h}}  \int_{\Sigma_t}
\frac{\kappa}{\sqrt{h}}
 \pi^{ij}\cG_{ijkl}\pi^{kl} -\frac{\sqrt{h}}{\kappa} \sR \right)
 \frac{1}{2}\sqrt{h}h^{ij}\bar\xi .
\end{align}
The algebra of gauge transformations is obtained as
\begin{equation}\label{ga.uADM}
 \delta_{\tilde\xi}\delta_{\tilde\psi}\varphi
-\delta_{\tilde\psi}\delta_{\tilde\xi}\varphi
  =\delta_{\left[\tilde\xi,\tilde\psi\right]}\varphi,
\end{equation}
where we find the algebra of gauge parameters as
\begin{align}\label{gpa.uADM}
 \left[\tilde\xi,\tilde\psi\right]^0&=-\left(
\overline{\xi^i\partial_i\bar\psi}
 -\overline{\partial_i\bar\xi\psi^i} \right),  \nonumber\\
 \left[\tilde\xi,\tilde\psi\right]^i&=-h^{ij}\left(
\bar\xi\partial_j\bar\psi-\partial_j\bar\xi\bar\psi \right)  -\left(
\xi^j\partial_j\psi^i-\partial_j\xi^i\psi^j \right).
\end{align}

\subsection{BRST Symmetry}

The BRST transformation is obtained as
\begin{subequations}
\begin{align}
 \delta_b h_{ij}&=\left( \frac{2\kappa}{\sqrt{h}}  \cG_{ijkl}\pi^{kl}
\bar{c}  +\partial_kh_{ij}c^k  +h_{ik}\partial_jc^k +h_{kj}\partial_ic^k
 \right) \theta,\label{brstA}\\
 \delta_b \pi^{ij}&=\left[ \frac{1}{2}h^{ij}\left(
\frac{\kappa}{\sqrt{h}}\pi^{kl}\cG_{klmn}\pi^{mn}
+\frac{\sqrt{h}}{\kappa}\sR  +\frac{\sqrt{h}}{\int_{\Sigma_t}\sqrt{h}}
\int_{\Sigma_t}\left(  \frac{\kappa}{\sqrt{h}}
\pi^{ij}\cG_{ijkl}\pi^{kl}
 -\frac{\sqrt{h}}{\kappa} \sR \right) \right)  \right. \nonumber \\
 &\quad-\left.\frac{\kappa}{\sqrt{h}}\left( 2\pi^{(i}_{\ \:k}\pi^{j)k}
 -\pi^{ij}h_{kl}\pi^{kl} \right)  -\frac{\sqrt{h}}{\kappa}\left(
 \sR^{ij} -D^iD^j +h^{ij}D^kD_k   \right) \right] \bar{c} \theta
\nonumber \\
 &\quad+\left( \partial_k\left( \pi^{ij}c^k \right)
 -\pi^{ik}\partial_kc^j
 -\pi^{kj}\partial_kc^i \right) \theta,  \\
 \delta_b \bar{c}&=-\frac{1}{2}\left[\tilde c,\tilde c\right]^0\theta
 =\overline{c^i\partial_i\bar{c}}\theta,  \\
 \delta_b c^i&=-\frac{1}{2}\left[\tilde c,\tilde c\right]^i\theta
 =\left(h^{ij}\bar{c}\partial_j\bar{c}+c^j\partial_jc^i\right)\theta, \\
 \delta_b \bar{c}^*&=\bar\eta\theta, \\
 \delta_b c_i^*&=\eta_i\theta, \\
 \delta_b \bar\eta&=0, \\
 \delta_b \eta_i&=0. \label{brstH}
\end{align}
\end{subequations}
The BRST transformation of metric $h_{ij}$ and the momentum $\pi^{ij}$
are obtained from their gauge transformations \eqref{gt.uADM.metric} and
\eqref{gt.uADM.pi}, respectively, by replacing the gauge with parameters
as $\bar\xi\rightarrow\bar{c}\theta$ and $\xi^i\rightarrow c^i\theta$.
The
transformation of the ghosts $\tilde c^\mu=(\bar{c},c^i)$ is
obtained from the algebra of gauge parameters \eqref{gpa.uADM} with the
same replacement. Since the generator $\bcH_T$ has a vanishing integral
over space, the ghosts $\bar{c}$, $\bar{c}^{*}$ and the field
$\bar\eta$ are average-free as well.

\subsection{Gauge fixing and ghost action }

As previously stated, the gauge generators in this unimodular setting
with fixed metric determinant are the average-free Hamiltonian and
super-momentum constraints, $\tcH_\mu=(\bcH_T,\cH_i)$, demanding that
one of the gauge conditions $\tilde\chi^\mu$ has to be average-free, so
that the number
of gauge conditions matches the number of generators exactly.
We choose it to be the zero-component, since the zero mode of the
super-Hamiltonian is a second class constraint, and hence we denote
$\tilde\chi^\mu=(\bar\chi^0,\chi^i)$.


\subsubsection{Unimodular Faddeev-Popov gauge}

The usual Faddeev-Popov (FP) gauge \cite{fad} is defined as
\begin{equation}\label{gauge.FP}
 \chi^{0}_{\mathrm{FP}}=\ln h-\Phi\approx0,\quad
 \chi^{1}_{\mathrm{FP}}=h_{23}\approx0,\quad
 \chi^{2}_{\mathrm{FP}}=h_{31}\approx0,\quad
 \chi^{3}_{\mathrm{FP}}=h_{12}\approx0,
\end{equation}
where $\ln h=\ln(\det h_{ij})$ and $\Phi$ is a fixed function. The
average-free component $\overline{\ln h}$ of $\ln h$ is not a scalar
density of any weight. Hence it is unclear which measure we should use
to integrate $\ln h$ over $\Sigma_t$. Here we treat $\ln h$ as a scalar,
so that
\begin{equation}
 \overline{\ln h}=\ln h-\frac{1}{\int_{\Sigma_t}\sqrt{h}}
\int_{\Sigma_t}\sqrt{h}\ln h.
\end{equation}
The unimodular FP gauge conditions are thus defined as
\begin{equation}\label{gauge.uFP}
 \bar\chi^{0}_{\mathrm{FP}}=\overline{\ln h}-\bar{\Phi}\approx0,\quad
 \chi^{i}_{\mathrm{FP}}=\frac{1}{2}d^{ijk}h_{jk}\approx0,
\end{equation}
where $\bar{\Phi}$ is a function with zero average,
$\int_{\Sigma_t}\sqrt{h}\bar{\Phi}=0$, and the last three conditions
$\chi^{i}_{\mathrm{FP}}$ ($i=1,2,3$) are identical to those in
\eqref{gauge.FP}, which impose the off-diagonal components of the metric
to vanish, but written with the help of a strictly positive permutation
symbol
\begin{equation}
 d^{ijk}=\left\{ \begin{aligned}
 1,& \quad \text{if the indices $ijk$ are any permutation of $123$,}\\
 0,& \quad \text{if any of the indices $ijk$ are  equal.}
 \end{aligned}\right.
\end{equation}
The BRST invariant gauge and ghost action in the unimodular FP gauge is
given by
\begin{align}\label{SuFP}
 S^\mathrm{FP}_{gf+gh}&=\int d^4x\bigg( -\sqrt{h}\bar\eta\left(
\overline{\ln h}-\bar{\Phi} \right)
-\frac{1}{2}\sqrt{h}\eta_id^{ijk}h_{jk} \nonumber\\
 &\quad-\sqrt{h}\bar{c}^{*}\left( \overline{\ln h} -\bar{\Phi}+2
 \right)\left( K\bar{c} +D_ic^i \right)- \sqrt{h}c^{*}_id^{ijk}\left(
K_{jk}+\frac{1}{2}h_{jk}K  \right)\bar{c}  \nonumber\\
&\quad  -\frac{1}{2}\sqrt{h}c^{*}_id^{ijk}\left( h_{jk}D_lc^l
 +\partial_lh_{jk}c^l +h_{jl}\partial_kc^l +h_{lk}\partial_jc^l \right)
\bigg).
\end{align}
It should be noted that \eqref{SuFP} is written in terms of the
extrinsic curvature $K_{ij}$ and not momentum $\pi^{ij}$. This is
because when the canonical momenta are integrated in the path integral,
the momentum $\pi^{ij}$ is expressed in terms of the metric variables as
\begin{equation}\label{pitoK}
 \pi^{ij}=\frac{\sqrt{h}}{\kappa}\cG^{ijkl}K_{kl}.
\end{equation}
Moreover, in obtaining \eqref{SuFP}, we used the fact that the
(average-free) ghost $\bar{c}^*$
has a vanishing average, so that for any time-dependent function $f(t)$
we obtain $ \int d^4x\bar{c}^*\sqrt{h}f(t)= 0$.

The path integral for the unimodular Faddeev-Popov gauge condition is
written as
\begin{align}
 Z^\mathrm{FP}_\mathrm{UG}&={\cal N}^{-1}\int\prod_x{\cal
D}g_{\mu\nu}{\cal D}\bar\eta  {\cal D}\eta_i {\cal D}\bar{c}^{*}{\cal
D}\bar{c}{\cal D}c^{*}_i  {\cal D}c^j g^{00}(-g)^{-\frac{3}{2}}
\nonumber \\
&\quad \times  \delta\left( \frac{\int_{\Sigma_t}\left(
\sqrt{-g}-\epsilon_0
\right)}{\left(-g^{00}\right)^{-\frac{1}{2}}\int_{\Sigma_t}\sqrt{h}}
\right)   \exp  \left( \frac{i}{\hbar}
\left[S_{EH}[g_{\mu\nu}]+S^{FP}_{gf+gh}\right]\right).\label{gen}
\end{align}

\subsubsection{Averaged metric determinant and spatial harmonic gauge}

To illustrate the analysis with further examples we consider now a mixed
unimodular condition. The first (average-free) gauge condition is chosen
to agree with the unimodular Faddeev-Popov gauge \eqref{gauge.uFP},
while the other conditions define harmonic coordinates on each spatial
hypersurface $\Sigma_t$:
\begin{equation}\label{gauge.DH}
 \bar\chi^{0}_{\mathrm{FP}}=\overline{\ln h}-\bar{\Phi}\approx0,\quad
 \chi^i_{\mathrm{H}}=\partial_j\left( \sqrt{h}h^{ij} \right)  \approx0.
\end{equation}
The BRST invariant gauge and ghost action reads
\begin{align}\label{SuDH}
 S^\mathrm{DH}_{gf+gh}&=\int d^4x\left( -\sqrt{h}\bar\eta\left(
\overline{\ln h} -\bar{\Phi} \right)  -\eta_i\partial_j\left(
\sqrt{h}h^{ij} \right)  -\sqrt{h}\bar{c}^{*}\left( \overline{\ln h}
-\bar{\Phi}+2 \right)\left(  K\bar{c} +D_ic^i \right) \right.
\nonumber\\
 &\quad+2c^{*}_i\partial_j\left[ \sqrt{h}\left( K^{ij}
-\frac{1}{2}h^{ij}K
\right)\bar{c} \right]
  -c^{*}_i\partial_j\partial_k\left( \sqrt{h}h^{ik} \right)c^j
\nonumber\\
  &\quad-c^{*}_i\partial_j\left( \sqrt{h}h^{ij} \right)\partial_kc^k
  +\left.c^{*}_i\partial_j\left( \sqrt{h}h^{jk} \right)\partial_kc^i
  +c^{*}_i\sqrt{h}h^{jk}\partial_j\partial_kc^i \right).
\end{align}
once again the momentum has been expressed in terms of metric variables
\eqref{pitoK}, and we denote $K^{ij}=h^{ik}h^{jl}K_{kl}$. Finally, the
path integral is given as
\begin{align}
 Z^\mathrm{DH}_\mathrm{UG}&={\cal N}^{-1}\int\prod_x{\cal
D}g_{\mu\nu}{\cal D}\bar\eta  {\cal D}\eta_i {\cal D}\bar{c}^{*}{\cal
D}\bar{c}{\cal D}c^{*}_i  {\cal D}c^j g^{00}(-g)^{-\frac{3}{2}}
\nonumber \\
& \quad \times \delta\left( \frac{\int_{\Sigma_t}\left(
\sqrt{-g}-\epsilon_0
 \right)}{\left(-g^{00}\right)^{-\frac{1}{2}}\int_{\Sigma_t}\sqrt{h}}
\right)
   \exp \left(
\frac{i}{\hbar}\left[S_{EH}[g_{\mu\nu}]+S^{DH}_{gf+gh}\right] \right) .
\end{align}

\subsubsection{Averaged metric trace and spatial harmonic gauge}

Another alternative gauge condition is proposed as: the first gauge
condition is chosen to be the average-free component of the trace of the
spatial metric, while the other conditions define harmonic coordinates
on each spatial hypersurface $\Sigma_t$:
\begin{equation}\label{gauge.TH}
 \bar\chi^{0}_{\mathrm{T}}=\overline{\mathrm{tr}(h_{ij})}
\approx0,\quad
 \chi^i_{\mathrm{H}}=\partial_j\left( \sqrt{h}h^{ij} \right)  \approx0,
\end{equation}
where
\begin{equation}
 \overline{\mathrm{tr}(h_{ij})}=\mathrm{tr}(h_{ij})
-\frac{1}{\int_{\Sigma_t}\sqrt{h}}
 \int_{\Sigma_t}\sqrt{h}\,\mathrm{tr}(h_{ij});\quad
\mathrm{tr}(h_{ij})=\sum_{i}h_{ii}.
\end{equation}
Hence, the BRST invariant gauge and ghost action in the trace gauge
condition is found to be
\begin{align}
 S^\mathrm{TH}_{gf+gh}&=\int d^4x\left( -\sqrt{h}\bar\eta
 \overline{\mathrm{tr}(h_{ij})}  -\eta_i\partial_j\left( \sqrt{h}h^{ij}
\right)
 -\sqrt{h}\bar{c}^{*}\overline{\mathrm{tr}(h_{ij})}\left(
 K\bar{c} +D_ic^i \right) \right.\nonumber\\
&\quad-\left. \sqrt{h}\bar{c}^*\sum_{i}\left( 2K_{ii}\bar{c}
+\partial_jh_{ii}c^j
 +2h_{ij}\partial_{i}c^j \right)  +2c^{*}_i\partial_j\left[
\sqrt{h}\left( K^{ij}
 -\frac{1}{2}h^{ij}K \right)\bar{c} \right]\right. \nonumber\\
  &\quad- \left. c^{*}_i\partial_j\partial_k\left( \sqrt{h}h^{ik}
\right)c^j
  -c^{*}_i\partial_j\left( \sqrt{h}h^{ij} \right)\partial_kc^k
  +  c^{*}_i\partial_j\left( \sqrt{h}h^{jk} \partial_kc^i \right)
\right).
\end{align}
Finally, the path integral is written as
\begin{align} \label{gen3}
 Z^\mathrm{TH}_\mathrm{UG}&={\cal N}^{-1}\int\prod_x{\cal
D}g_{\mu\nu}{\cal D}\bar\eta  {\cal D}\eta_i {\cal D}\bar{c}^{*}{\cal
D}\bar{c}{\cal D}c^{*}_i  {\cal D}c^j g^{00}(-g)^{-\frac{3}{2}}
\nonumber  \\
& \quad \times \delta\left( \frac{\int_{\Sigma_t}\left(
\sqrt{-g}-\epsilon_0
\right)}{\left(-g^{00}\right)^{-\frac{1}{2}}\int_{\Sigma_t}\sqrt{h}}
\right)  \exp
\left(\frac{i}{\hbar}\left[S_{EH}[g_{\mu\nu}]+S^\mathrm{TH}_{gf+gh}
\right]\right).
\end{align}

Before concluding this section, we mention a problem that can appear in
the present theory if one uses a (average-free) gauge condition that
involves the canonical momentum $\pi^{ij}$. In particular, adapting the
usual Dirac gauge conditions to the present unimodular theory with fixed
metric determinant involves a problem which is discussed in
Appendix~\ref{appA}.

With this section we conclude the first part of our analysis by
discussing the BRST invariant approach for the DUG and UG theory. We
have determined the BRST invariant path integral for both
theories for a set of gauge conditions. We now proceed further and
extend the previous study by establishing connections between transition
amplitude in different gauges. To achieve this goal, we shall first
introduce the finite field-dependent BRST transformations.

\section{The Generalized BRST transformation}
\label{sec3}

In this section, we illustrate the FFBRST (generalized BRST) formulation
\cite{sdj}
for the unimodular gravity theory with full diffeomorphism invariance in
an elegant way.  For that matter, we first write the BRST transformation
for all the fields of the theory,
Eqs.\eqref{BRSTtransa}-\eqref{BRSTtransd},
denoted collectively as $\phi_a(x)\equiv \phi(x)$, as follows:
\begin{align}
  \phi(x)\longrightarrow \phi^\prime(x)=\phi (x)+s_b \phi(x)\ \theta,
\end{align}
where $s_b\phi$ is the Slavnov variation of the field $\phi(x)$ and
$\theta$ is a Grassmann \emph{global} parameter.

To generalize the BRST symmetry, we first make all the fields $\phi(x)$
depend on a continuous parameter $\kappa$ ($0\leq \kappa\leq1$) in such
a way that the conditions $\phi(x, \kappa =0) \equiv \phi (x)$ and
$\phi(x, \kappa =1) \equiv \phi^\prime (x)=\phi (x)+ s_b\phi(x) \theta
[\phi]$ stand for the original field and the FFBRST transformed field,
respectively, where $\theta[\phi]$ is now a (functional) finite
\emph{field-dependent} parameter. Moreover, the FFBRST transformation is
justified by the following infinitesimal field-dependent BRST
transformation:
\begin{align}
  \frac{dg_{\mu\nu}(x, \kappa)}{d\kappa}&=\left( \partial_\rho
g_{\mu\nu}c^\rho
 +g_{\mu\rho}\partial_\nu c^\rho+g_{\rho\nu}\partial_\mu c^\rho \right)
 \theta^\prime[\phi (\kappa)],\nonumber\\
 \frac{dc^\mu(x, \kappa)}{d\kappa} &=-c^\nu\partial_\nu c^\mu
\theta^\prime[\phi (\kappa)],\nonumber\\
  \frac{d \bar c_\mu(x, \kappa)}{d\kappa} &=\eta_\mu\theta^\prime[\phi
(\kappa)],\nonumber\\
  \frac{d \eta_\mu (x, \kappa)}{d\kappa} &=0.
\end{align}
Integrating these equations with respect to $\kappa$, we find the
following field-dependent transformations
\begin{align}
 g_{\mu\nu}(x, \kappa) &= g_{\mu\nu}(x, 0)+\left( \partial_\rho
g_{\mu\nu}c^\rho
 +g_{\mu\rho}\partial_\nu c^\rho+g_{\rho\nu}\partial_\mu c^\rho \right)
 \theta [\phi (\kappa)],\nonumber\\
 c^\mu(x, \kappa)  &= c^\mu(x, 0)-c^\nu\partial_\nu c^\mu \theta [\phi
(\kappa)],\nonumber\\
\bar c_\mu(x, \kappa)   &=\bar c_\mu(x, 0)  +\eta_\mu\theta [\phi
(\kappa)],\nonumber\\
 \eta_\mu (x, \kappa)  &=0,
\end{align}
where we have $ \theta[\phi(\kappa)]$ as a functional of the fields
$\phi(x,\kappa)$ \cite{sdj}
\begin{align}
\label{theta}
 \theta[\phi(\kappa)]&=\int_0^\kappa d\kappa\
\theta^\prime[\phi(\kappa)],\nonumber\\
 &=\theta'[\phi(0)]\frac{\exp{\left(  \kappa
\frac{\delta\theta'}{\delta\phi}s_b\phi\right)}-1 }{
\frac{\delta\theta'}{\delta\phi}s_b\phi}.
\end{align}
At the boundary value of $\kappa$, i.e. $\kappa =1$, these expressions
yield to the FFBRST transformations,
\begin{align}
 \delta_b g_{\mu\nu}(x) &=  \left( \partial_\rho g_{\mu\nu}c^\rho
+g_{\mu\rho}\partial_\nu c^\rho+g_{\rho\nu}\partial_\mu c^\rho \right)
\theta [\phi (1)],\nonumber\\
 \delta_b  c^\mu(x)  &=  -c^\nu\partial_\nu c^\mu \theta
[\phi(1)],\nonumber\\
 \delta_b \bar c_\mu(x )   &=  \eta_\mu\theta [\phi
(1)],\nonumber\\
 \delta_b  \eta_\mu (x )  &=0,\label{ffb}
\end{align}
where finite field-dependent parameter reads $\theta [\phi(1)]=\theta
[\phi(\kappa)]_{\kappa=1}$.

Here we notice that the resulting FFBRST transformations with
field-dependent parameter \eqref{ffb} are a symmetry of the effective
action. However, the path integral measure changes non-trivially under
these leading thus to a non-trivial Jacobian. Hence, it is necessary
derive the explicit expression of the Jacobian for the functional
measure under the FFBRST transformations for an arbitrary $\theta$
parameter.

\subsection{Jacobian for field-dependent BRST transformation}

To compute the Jacobian we first define the path integral for
unimodular gravity theory in a general gauge as follows,
\begin{align}
 Z=\int {\cal D}\Phi\ e^{ \left(\frac{i}{\hbar}S_{EH}[\phi]+
S_{gf+gh}[\phi]\right)},\label{zen}
\end{align}
where ${\cal D}\Phi$ is the (BRST) covariant functional measure and
$S_{gf+gh}[\phi]$ refers to the general gauge-fixing and ghost part of
the effective action. In order to determine the Jacobian expression for
the
functional measure under the FFBRST transformations, we write \cite{sdj}
\begin{align}
 {\cal D}\Phi (\kappa) = J(\kappa) {\cal D}\Phi (\kappa) = J(\kappa
+d\kappa) {\cal D}\Phi (\kappa +d\kappa).
\end{align}
Since the transformation from $\phi(\kappa)$ to $\phi(\kappa+d\kappa)$
is viewed as an infinitesimal one, this can further be written as
\cite{sdj}
\begin{align}
 \frac{J(\kappa)}{J(\kappa +d\kappa) }  = \sum_\phi\pm
\frac{{\delta}\phi (\kappa +d\kappa)}{{\delta}\phi (\kappa)},
\end{align}
where $\pm$ sign is used for bosonic and fermionic fields, respectively.
Now, upon Taylor expansion the above expression yields
\begin{align}
 1-\frac{1}{J}\frac{dJ}{d\kappa} d\kappa =1+ d\kappa\int d^4x
\sum_\phi\pm s_b\phi(x,\kappa)
\frac{\delta\theta^\prime[\phi( \kappa)]}{\delta\phi( \kappa)},
\end{align}
which further simplifies to
\begin{align}
 \frac{d\ln J[\phi]}{d\kappa} =-\int d^4x \sum_\phi\pm s_b\phi(x,\kappa)
\frac{\delta\theta^\prime[\phi(\kappa)]}{\delta\phi( \kappa)}.
\end{align}
We now  perform the integration over $\kappa$ (after Taylor expansion)
with an appropriate limit,
to get the following:
\begin{align}
 \ln J [\phi]  =&-\int_0^1 d\kappa\int d^4x \sum_\phi\pm
s_b\phi(x,\kappa)
\frac{\delta\theta^\prime[\phi( \kappa)]}{\delta\phi( \kappa)},
\nonumber\\
=&- \left(\int d^4x \sum_\phi\pm s_b\phi(x)
\frac{\delta\theta^\prime[\phi ]}{\delta\phi }\right).
\end{align}
This result leads to the final expression for the Jacobian generated
from a variation of the functional measure under FFBRST
transformations with an arbitrary parameter
\begin{align}
 J[\phi]   =   \exp\left(-\int d^4x \sum_\phi\pm s_b\phi(x)
\frac{\delta\theta^\prime[\phi]}{\delta\phi}\right).\label{J}
\end{align}
We remark here that this  expression of Jacobian is rather elegant than
one
originally derived  in \cite{sdj}. Since the Jacobian obtained here
depends explicitly on the parameter $\theta'$.

Now, with the expression \eqref{J} for the Jacobian
(generated by FFBRST transformation) we find that the path integral
\eqref{zen} changes as
 \begin{align}
 \int {\cal D}\Phi^\prime \ e^{\left(\frac{i}{\hbar}S_{EH}[\phi^\prime]+
S_{gf+gh}[\phi^\prime]\right) }&=\int J[\phi] {\cal D}\Phi \ e^{
\left(\frac{i}{\hbar}S_{EH}[\phi]+ S_{gf+gh}[\phi]\right)}\nonumber\\
 &=\int {\cal D}\Phi \ e^{ \left(\frac{i}{\hbar} S_{EH}[\phi]+
S_{gf+gh}[\phi] -\int d^4x \left(\sum_\phi\pm s_b\phi
\frac{\delta\theta^\prime}{\delta\phi }\right)\right)}.
 \end{align}
This is the FFBRST transformed path integral of the
unimodular gravity theories (both DUG and UG) with an extended
action, where the gauge fixing and ghosts actions are modified by the
Jacobian. We emphasize that the form of the functional parameter
$\theta^\prime$ should be chosen so that the Jacobian \eqref{J} does
not produce any physical change in the quantum theory.
Otherwise, one could choose $\theta^\prime$ so that the physical content
of the quantum theory is modified, e.g. producing new vertices and/or
propagating modes, which would not be a symmetry transformation.
For this matter we emphasize that we
consider in our analysis only the path integral of the vacuum transition
amplitude.
We shall now illustrate this result by establishing the connection
between different gauges of the two presented versions of unimodular
gravity.


\subsection{Connection of different gauges in fully
diffeomorphism-invariant theory}

In this section we study the connection of various important gauges of
the fully diffeomorphism-invariant unimodular gravity (as stated in
section~\ref{sec1}). In particular, notice that these are well-defined
gauges, since then there should be no physical change in the quantum
theory.
We will show the connection between the following
gauges: (i) harmonic and synchronous gauges, (ii) axial and harmonic
gauges, (iii) harmonic and Lorentz gauges, and, at last, (iv) Lorentz
and
synchronous gauges.

\subsubsection{Harmonic to  synchronous gauge}

For this analysis, we follow the standard procedure as discussed above.
We first construct the infinitesimal version of the functional parameter
\eqref{ffb} as follows
\begin{equation}
\theta^\prime[\phi]=-\int d^4x \left[-\bar c_\mu \partial_\nu \hat
g^{\mu\nu}
+\sqrt{-g}\bar c^0 (g_{00}+1)+\sqrt{-g} \bar c^ig_{0i}
\right].\label{th}
\end{equation}
The advantage of constructing an infinitesimal version is that with such
parameter the Jacobian can be computed directly from \eqref{J}. Thus,
the Jacobian expression for this choice of parameter \eqref{th} is
\begin{align}\label{J1}
  J[\phi]   &=  \exp\bigg[ \int
d^4x\bigg(\eta_\mu\partial_\nu\hg^{\mu\nu}
 +\bar{c}_\mu\left[  \partial_\nu \left(
\partial_\rho(\hg^{\mu\nu}c^\rho)
 -\hg^{\mu\rho}\partial_\rho c^\nu  -\hg^{\rho\nu}\partial_\rho c^\mu
\right)\right]  \nonumber\\
&\quad-\sqrt{-g}\eta^0(g_{00}+1)-\sqrt{-g}\eta^ig_{0i}
 -\sqrt{-g}\bar{c}^0\nabla_\mu c^\mu \nonumber\\
&\quad- \sqrt{-g}\bar{c}^\mu\left( g_{0\mu}\nabla_\nu c^\nu
+\partial_\nu
g_{0\mu} c^\nu +g_{0\nu}\partial_\mu c^\nu  +g_{\mu\nu}\partial_0 c^\nu
\right)  \bigg) \bigg].
\end{align}
With this Jacobian the generating functional in harmonic gauge
\eqref{PIH} changes to
 \begin{align}
 {\cal N}^{-1}\int {\cal D}\Phi ^\prime \
e^{i\left(S_{EH}[\phi^\prime]+ S^H_{gf+gh}[\phi^\prime]\right)
}&={\cal N}^{-1}\int J[\phi] {\cal D}\Phi \  e^{i\left(S_{EH}[\phi]+
S^H_{gf+gh}[\phi]\right)}\nonumber\\
 &={\cal N}^{-1}\int {\cal D}\Phi
\  e^{i( S_{EH}[\phi]+
S^S_{gf+gh}[\phi])}\nonumber\\
 & = Z^S_\mathrm{DUG},
 \end{align}
 which is nothing but the transition amplitude in synchronous gauge
\eqref{fd}.  Here $\phi^\prime$ and $\phi$ denote, respectively, the
transformed and
generic fields of the DUG theory. The invariant functional measure for
DUG is defined as
${\cal D}\Phi\equiv\prod_x{\cal D}g_{\mu\nu} {\cal D}\eta_\mu{\cal
D}\bar{c}_\mu {\cal D}c^\nu  g^{00}(-g)^{-\frac{3}{2}}$. Thus the FFBRST
transformation with
parameter \eqref{th} establishes the connection between harmonic and
synchronous gauges, Eqs.\eqref{gauge.H} and \eqref{gauge.S},
respectively, for fully diffeomorphism-invariant unimodular gravity
theory.

 \subsubsection{Axial to harmonic gauge}

 To relate axial and harmonic gauges, Eqs.\eqref{gauge.A} and
\eqref{gauge.H}, respectively, we consider the following infinitesimal
field-dependent parameter
\begin{equation}
\theta^\prime[\phi]=-\int d^4x \left[  -a_{(\mu}\eta_{\nu)}\hat
g^{\mu\nu}+
\bar c_\mu\partial_\nu \hat g^{\mu\nu}\right].\label{gw}
\end{equation}
The Jacobian for functional measure under FFBRST transformation is
calculated by
\begin{align} \label{J2}
  J[\phi] &= \exp\bigg[ \int d^4x\bigg( a_{(\mu}\eta_{\nu)}\hg^{\mu\nu}
 +a_{(\mu}\bar{c}_{\nu)}\left[ \partial_\rho(\hg^{\mu\nu}c^\rho)
 -\hg^{\mu\rho}\partial_\rho c^\nu  -\hg^{\rho\nu}\partial_\rho c^\mu
\right] \nonumber\\
&  \quad -\eta_\mu\partial_\nu\hg^{\mu\nu}  +\bar{c}_\mu\left[
-\partial_\nu
\left(
\partial_\rho(\hg^{\mu\nu}c^\rho)  -\hg^{\mu\rho}\partial_\rho c^\nu
-\hg^{\rho\nu}\partial_\rho c^\mu \right) \right] \bigg)\bigg].
 \end{align}
Now substituting this Jacobian \eqref{J2} into the expression of path
integral measure in axial gauge \eqref{PIA} as follows
 \begin{align}
 {\cal N}^{-1}\int {\cal D}\Phi ^\prime
\  e^{i\left(S_{EH}[\phi^\prime]+ S^A_{gf+gh}[\phi^\prime]\right)
}&={\cal N}^{-1}\int J[\phi] {\cal D}\Phi
\  e^{i\left(S_{EH}[\phi]+ S^A_{gf+gh}[\phi]\right)}\nonumber\\
 &={\cal N}^{-1}\int {\cal D}\Phi
\  e^{i( S_{EH}[\phi]+S^H_{gf+gh}[\phi])}\nonumber\\
 & = Z^H_\mathrm{DUG},
 \end{align}
 and we thus get the expression of path integral in harmonic
gauge \eqref{PIH}.
 Therefore,  FFBRST transformation, generated with the parameter
\eqref{gw}, connects the axial and harmonic gauges of the theory.

 Here we remark that the same value of Jacobian given in
\eqref{J2} when replaced into the expression of the transition amplitude
in Lorentz
gauge \eqref{fd} gives the transition amplitude in planar gauge
\eqref{genp}. Thus, the FFBRST transformation with parameter \eqref{gw}
also connects the Lorentz gauge \eqref{gauge.L} to planar gauge
\eqref{gauge.P}.

\subsubsection{Harmonic to Lorentz gauge}

To establish the connection of the harmonic gauge to Lorentz gauge,
Eqs.\eqref{gauge.H} and \eqref{gauge.L}, respectively, we determine the
infinitesimal functional parameter as follows
\begin{equation}
\theta^\prime[\phi]=-\int d^4x \left[ \bar c_\mu \frac{\alpha}{2}\hat
g_R^{\mu\nu} \eta_\nu\right].\label{ht}
\end{equation}
Utilizing this parameter the Jacobian for path integral measure is
calculated by
\begin{align} \label{J3}
  J[\phi] = \exp\left[ \int d^4x\left(
-\frac{\alpha}{2}\hg^{\mu\nu}_\mathrm{R}\eta_\mu\eta_\nu \right)
\right] .
 \end{align}
 This value for the Jacobian when inserted into the transition amplitude
changes the theory from the harmonic gauge \eqref{PIH} into the one in
the Lorentz gauge \eqref{fd} as follows
 \begin{align}
 {\cal N}^{-1}\int {\cal D}\Phi ^\prime \
e^{i\left(S_{EH}[\phi^\prime]+ S^H_{gf+gh}[\phi^\prime]\right) }&={\cal
N}^{-1}\int J[\phi] {\cal D}\phi  \
e^{i\left(S_{EH}[\phi]+ S^H_{gf+gh}[\phi]\right)}\nonumber\\
 &={\cal N}^{-1}\int {\cal D}\phi   \  e^{i(
S_{EH}[\phi]+
S^L_{gf+gh}[\phi])}\nonumber\\
 & = Z^L_\mathrm{DUG}.
 \end{align}

 Here we emphasize that the Jacobian expression \eqref{J3} is also
responsible
to connect the axial gauge \eqref{gauge.A} to planar gauge
\eqref{gauge.P}. Thus the path integral for DUG in axial gauge
\eqref{PIA} under FFBRST transformation with parameter \eqref{ht}
switches to the transition amplitude in planar gauge \eqref{genp}.

\subsubsection{Lorentz to  synchronous gauge}

Finally, we determine the connection between Lorentz gauge and
synchronous gauge, Eqs.\eqref{gauge.L} and \eqref{gauge.S},
respectively. For this purpose we construct  the functional parameter as
follows
\begin{equation}
\theta^\prime[\phi]=-\int d^4x \left[ -\bar c_\mu \frac{\alpha}{2}\hat
g_R^{\mu\nu} \eta_\nu -\bar c_\mu \partial_\nu \hat g^{\mu\nu}
+\sqrt{-g}\bar c^0
(g_{00}+1)+\sqrt{-g} \bar c^ig_{0i}\right].
\end{equation}
The corresponding Jacobian is found to read
\begin{align}\label{J4}
  J[\phi]   &=  \exp\bigg[ \int d^4x\bigg(
\frac{\alpha}{2}\hg^{\mu\nu}_\mathrm{R}\eta_\mu\eta_\nu   +
\eta_\mu\partial_\nu\hg^{\mu\nu}  +\bar{c}_\mu\left[  \partial_\nu
\left( \partial_\rho(\hg^{\mu\nu}c^\rho)  -\hg^{\mu\rho}\partial_\rho
c^\nu  -\hg^{\rho\nu}\partial_\rho c^\mu \right)\right]  \nonumber\\
 &\quad -\sqrt{-g}\eta^0(g_{00}+1)-\sqrt{-g}\eta^ig_{0i}
 -\sqrt{-g}\bar{c}^0\nabla_\mu c^\mu \nonumber\\
 & \quad -  \sqrt{-g}\bar{c}^\mu\left( g_{0\mu}\nabla_\nu c^\nu
+\partial_\nu
g_{0\mu} c^\nu +g_{0\nu}\partial_\mu c^\nu  +g_{\mu\nu}\partial_0 c^\nu
\right)  \bigg) \bigg].
\end{align}
Substituting this value \eqref{J4} into the generating functional in
Lorentz gauge \eqref{fd} we get
\begin{align}
{\cal N}^{-1}\int {\cal D}\Phi ^\prime  \
e^{i\left(S_{EH}[\phi^\prime]+ S^\alpha_{gf+gh}[\phi^\prime ]\right)
}&={\cal N}^{-1}\int J[\phi] {\cal D}\Phi \
e^{i\left(S_{EH}[\phi]+ S^\alpha_{gf+gh}[\phi]\right)}\nonumber\\
&={\cal N}^{-1}\int {\cal D}\Phi  \  e^{i(
S_{EH}[\phi]+
S^S_{gf+gh}[\phi])}\nonumber\\
& = Z^S_\mathrm{DUG}.
\end{align}
This establishes a connection between the path integral on
Lorentz gauge \eqref{fd} and synchronous gauge \eqref{PIS}.

Hence we concluded this subsection of analysis of FFBRST equivalence by
establishing relations among different and relevant gauge conditions of
fully-diffeomorphism invariant theory of unimodular gravity. Next we
will perform a similar analysis but now for unimodular gravity with
fixed metric determinant.


\subsection{Connection of different gauges in unimodular gravity with
fixed metric determinant}

In this subsection we analyse the connection of different gauges of
unimodular gravity with fixed metric determinant. Following the results
from section~\ref{sec2}, the FFBRST transformation for unimodular
gravity with fixed metric determinant are determined by the replacement
of the parameter $\theta \rightarrow \theta[\phi] $ into the
Eqs.\eqref{brstA}-\eqref{brstH}.

With these results we will show the following mapping: (i) unimodular
Faddeev-Popov to averaged metric determinant and spatial harmonic
gauges, (ii) unimodular Faddeev-Popov to averaged metric trace and
spatial harmonic gauges, and, finally, (iii) averaged metric
determinant
to averaged metric trace gauges.


\subsubsection{Unimodular Faddeev-Popov to averaged metric determinant
and spatial harmonic gauges}

In order to map the unimodular Faddeev-Popov and averaged metric
determinant and spatial harmonic gauges, Eqs.\eqref{gauge.uFP} and
\eqref{gauge.DH}, respectively, we define the infinitesimal
field-dependent parameter as follows
\begin{equation}
\theta^\prime[\phi]=-\int d^4x \left[ -\frac{1}{2}c^*_i \sqrt{h}
d^{ijk}h_{jk}
+c^*_i\partial_j(\sqrt{h} h^{ij})\right].
\end{equation}
Now  with the help of expression \eqref{J} we compute the respective
Jacobian corresponding to this parameter
\begin{align}\label{J5}
  J[\phi]   &=  \exp\bigg[ \int d^4x\bigg(
\frac{1}{2}\sqrt{h}\eta_id^{ijk}h_{jk} +\sqrt{h}c^{*}_id^{ijk}\left(
K_{jk}+\frac{1}{2}h_{jk}K  \right)\bar{c} \nonumber\\
&\quad+   \frac{1}{2}\sqrt{h}c^{*}_id^{ijk}\left( h_{jk}D_lc^l
+\partial_lh_{jk}c^l +h_{jl}\partial_kc^l +h_{lk}\partial_jc^l \right)
 -\eta_i\partial_j\left( \sqrt{h}h^{ij} \right) \nonumber\\
 &\quad+ 2c^{*}_i\partial_j\left[ \sqrt{h}\left( K^{ij}
-\frac{1}{2}h^{ij}K
\right)\bar{c} \right]
  -c^{*}_i\partial_j\partial_k\left( \sqrt{h}h^{ik} \right)c^j
-c^{*}_i\partial_j\left( \sqrt{h}h^{ij} \right)\partial_kc^k
\nonumber\\
  &\quad+   c^{*}_i\partial_j\left( \sqrt{h}h^{jk} \right) \partial_kc^i
+
c^{*}_i \sqrt{h}h^{jk} \partial_j\partial_k c^i \bigg) \bigg].
\end{align}
With this result for the Jacobian \eqref{J5} the transition amplitude
for unimodular gravity with fixed metric determinant in Faddeev-Popov
gauge \eqref{gen} changes as
\begin{align}
{\cal N}^{-1}\int {\cal D}\Phi ^\prime \ e^{\frac{i}{\hbar}\left(S_{EH}[\phi^\prime]+S^{FP}_{gf+gh}\right)} &={\cal N}^{-1}\int J[\phi] {\cal D}\Phi\  e^{\frac{i}{\hbar}\left(S_{EH}[\phi]+S^{FP}_{gf+gh}\right)}
\nonumber\\
 &={\cal N}^{-1}\int  {\cal D}\Phi  \  e^{\frac{i}{\hbar}\left(S_{EH}[\phi]+S^{DH}_{gf+gh}\right)}
\nonumber\\
& = Z^{DH}_\mathrm{UG},
\end{align}
which is exactly the expression for the path integral in
averaged metric determinant and spatial harmonic gauge. Here the
explicit expression for the invariant functional measure is now given as,  ${\cal D}\Phi \equiv\prod_x{\cal D}g_{\mu\nu}{\cal D}\bar\eta  {\cal D}\eta_i {\cal D}\bar{c}^{*}{\cal D}\bar{c}{\cal D}c^{*}_i  {\cal D}c^j g^{00}(-g)^{-\frac{3}{2}}\delta\left( \frac{\int_{\Sigma_t}\left( \sqrt{-g}-\epsilon_0 \right)}{\left(-g^{00}\right)^{-\frac{1}{2}}\int_{\Sigma_t}\sqrt{h}} \right)$ .


\subsubsection{Unimodular Faddeev-Popov to averaged metric trace and spatial harmonic gauges}

To connect the unimodular Faddeev-Popov gauge \eqref{gauge.FP} to
averaged metric trace and spatial harmonic gauge \eqref{gauge.TH} we
derive the transformation functional parameter as follows
\begin{equation}
\theta^\prime[\phi]=-\int d^4x \left[ -\bar c^* \sqrt{h}(\overline{\ln
h}-\bar{\Phi}-\overline{\mathrm{tr}(h_{ij})}) +c^*_i \sqrt{h}\left(
-\frac{1}{2}d^{ijk} h_{jk}- \partial_j(\sqrt{h}
h^{ij})\right)\right].\label{the}
\end{equation}
With this parameter the Jacobian of functional measure is calculated by
\begin{align} \label{J6}
  J[\phi]   &=  \exp\bigg[ \int d^4x\bigg( \sqrt{h}\bar\eta\left(
\overline{\ln h}-\bar{\Phi} \right)
+\frac{1}{2}\sqrt{h}\eta_id^{ijk}h_{jk}  +\sqrt{h}\bar{c}^{*}\left(
\overline{\ln h} -\bar{\Phi}+2  \right)\left( K\bar{c} +D_ic^i \right)
\nonumber\\
 &\quad+\sqrt{h}c^{*}_id^{ijk}\left( K_{jk}+\frac{1}{2}h_{jk}K
\right)\bar{c}  +\frac{1}{2}\sqrt{h}c^{*}_id^{ijk}\left( h_{jk}D_lc^l
+\partial_lh_{jk}c^l +h_{jl}\partial_kc^l +h_{lk}\partial_jc^l \right)
\nonumber\\
  & \quad- \sqrt{h}\bar\eta  \overline{\mathrm{tr}(h_{ij})}
-\eta_i\partial_j\left( \sqrt{h}h^{ij} \right)
-\sqrt{h}\bar{c}^{*}\overline{\mathrm{tr}(h_{ij})}\left( K\bar{c}
+D_ic^i \right) \nonumber\\
 & \quad- \sqrt{h}\bar{c}^*\sum_{i}\left( 2K_{ii}\bar{c} +\partial_jh_{ii}c^j
 +2h_{ij}\partial_{i}c^j \right)  +2c^{*}_i\partial_j\left[
\sqrt{h}\left( K^{ij}
 -\frac{1}{2}h^{ij}K \right)\bar{c} \right]  \nonumber\\
  &\quad-   c^{*}_i\partial_j\partial_k\left( \sqrt{h}h^{ik} \right)c^j
-c^{*}_i\partial_j\left( \sqrt{h}h^{ij} \right)\partial_kc^k  +
c^{*}_i\partial_j\left( \sqrt{h}h^{jk}
\right)\partial_kc^i  \nonumber\\
    &\quad+  c^{*}_i\sqrt{h}h^{jk}\partial_j\partial_kc^i   \bigg)
\bigg].
\end{align}
This Jacobian \eqref{J6} amounts the following change into the
expression of transition amplitude \eqref{gen}
\begin{align}
{\cal N}^{-1}\int {\cal D}\Phi^\prime \
e^{\frac{i}{\hbar}\left(S_{EH}[\phi^\prime]+S^{FP}_{gf+gh}\right)}
&={\cal N}^{-1}\int J[\phi] {\cal D}\Phi\  e^{\frac{i}{\hbar}\left(S_{EH}[\phi]+S^{FP}_{gf+gh}\right)}
\nonumber\\
 &={\cal N}^{-1}\int  {\cal D}\Phi\  e^{\frac{i}{\hbar}\left(S_{EH}[\phi]+S^{TH}_{gf+gh}\right)}
\nonumber\\
& = Z^{TH}_\mathrm{UG}.
\end{align}
This relation assures the connection (under FFBRST transformation)
between path integrals in the unimodular Faddeev-Popov and
averaged metric trace and spatial harmonic gauges, Eqs.\eqref{gen} and
\eqref{gen3}, respectively.


\subsubsection{Averaged metric determinant to averaged metric trace gauge}

Finally, we establish a connection between averaged metric determinant to averaged metric trace gauges, Eqs.\eqref{gauge.DH} and \eqref{gauge.TH},
respectively. For this purpose, we construct the following infinitesimal
field-dependent parameter:
\begin{equation}
\theta^\prime[\phi]=-\int d^4x \left[ -\bar c^* \sqrt{h}(\overline{\ln
h}-\bar{\Phi}-\overline{\mathrm{tr}(h_{ij})}) \right].\label{the1}
\end{equation}
The Jacobian expression \eqref{J} together with \eqref{the1} yields
\begin{align}\label{J7}
  J[\phi]   &=  \exp\bigg[ \int d^4x\bigg( \sqrt{h}\bar\eta\left(
\overline{\ln h} -\bar{\Phi} \right)   +\sqrt{h}\bar{c}^{*}\left(
\overline{\ln h} -\bar{\Phi}+2
\right)\left(  K\bar{c} +D_ic^i \right)  \nonumber\\
  &\quad- \sqrt{h}\bar\eta
 \overline{\mathrm{tr}(h_{ij})} -   \sqrt{h}\bar{c}^{*}\overline{\mathrm{tr}(h_{ij})}\left( K\bar{c}
+D_ic^i \right)\nonumber\\
  &\quad   -  \sqrt{h}\bar{c}^*\sum_{i}\left( 2K_{ii}\bar{c} +\partial_jh_{ii}c^j
 +2h_{ij}\partial_{i}c^j \right)   \bigg) \bigg].
\end{align}
It can directly be seen that this Jacobian \eqref{J7} is responsible
for the connection of  averaged metric determinant gauge to averaged metric
trace gauge as follows
\begin{align}
 {\cal N}^{-1}\int {\cal D}\Phi^\prime  \
e^{\frac{i}{\hbar}\left(S_{EH}[\phi^\prime]+S^{DH}_{gf+gh}\right)}
&={\cal N}^{-1}\int J[\phi] {\cal D}\Phi\  e^{\frac{i}{\hbar}\left(S_{EH}[\phi]+S^{DH}_{gf+gh}\right)} \nonumber\\
 &={\cal N}^{-1}\int  {\cal D} \Phi \ e^{\frac{i}{\hbar}\left(S_{EH}[\phi]+S^{TH}_{gf+gh}\right)} \nonumber\\
& = Z^{TH}_\mathrm{UG}.
\end{align}
Thus we conclude this subsection where we have explicitly presented a
detailed analysis concerning the FFBRST transformation equivalence (with
specific choices for the parameters) relating various gauges of the
unimodular gravity with fixed metric determinant.

%

\section{Concluding Remarks}
\label{sec4}

As we know a gauge invariant theory can not be quantized correctly
without fixing the gauge properly. Being a gauge theory, we have
discussed the implementation of various gauge conditions for two version
of the unimodular gravitational theory, fully diffeomorphism-invariant
unimodular gravity and unimodular gravity with fixed metric determinant.
We have further incorporated these gauges together with ghost terms at
quantum level by defining the respective path integral. Further on, we
derived the nilpotent BRST symmetry for the effective action as well as
for the transition amplitude.

In particular, it should be noted that, in the fully diffeomorphism invariant unimodular gravity
\cite{Bufalo:2015wda}, after the auxiliary variables of action
\eqref{SDUG} have been integrated out, the gauge symmetry of the path
integral \eqref{ZDUG} is the same as that of GR. Therefore the
formulation of gauge conditions and the associated gauge fixing and
ghost actions can be achieved in a familiar way. We obtained the gauge
fixing and ghost action for several relevant gauges in
section~\ref{sec1}. The results can be applied to both (DUG) unimodular
gravity and GR due to the similar gauge symmetry.

Furthermore, we have formulated three possible gauges for unimodular
gravity theory with fixed metric determinant \eqref{UG} in
section~\ref{sec2}. In this case, gauge fixing is more involving since
the gauge symmetry of the theory has been restricted, so that the
unimodular condition remains gauge invariant. Consequently, the integral
of the Hamiltonian constraint over space is not a generator of a gauge
transformation, and hence the integral of one of the gauge conditions must vanish, and the corresponding ghost and antighost fields are
average-free as well (see \cite{Bufalo:2015wda} for a detailed analysis). In some cases, this restricted gauge structure may complicate the
formulation of gauge conditions and BRST invariant actions, in particular, if the
chosen gauge conditions involve the canonical momentum conjugate to the
induced metric on the spatial hypersurface; an example of this problem
is discussed in Appendix~\ref{appA}.

The BRST symmetry of these theories has been further extended by making
the transformation parameter finite and field-dependent. We have shown
that the FFBRST transformation of the Jacobian of the invariant functional
measure, with specific choices for the transformation parameter,
connects various gauges of both given unimodular theories of gravity.
This establishes a way to consistently relate several path integral expressions when defined in different gauge conditions.
However, we should emphasize that we are using the FFBRST
transformation only for connecting different well-defined gauges, since
then there should be no physical change in the quantum theory.
Thus FFBRST formulation discussed here could be useful in comparing
results in two gauges for unimodular gravity theories.

\subsection*{Acknowledgments}
M.O. thankfully acknowledges support from the Emil Aaltonen Foundation.
R.B. acknowledges FAPESP for full support, Project No. 2011/20653-3.


\appendix
\section{Unimodular Dirac gauge}
\label{appA}

In order to justify the absence of the Dirac gauge in our analysis of
unimodular gravity with fixed metric determinant, we highlight a problem
in the formulation of a gauge condition that depends on the canonical
momentum $\pi^{ij}$ conjugate to the induced metric $h_{ij}$. The
Dirac gauge could be defined in the unimodular setting as
\begin{equation}\label{gauge.uDirac}
\bar\chi^{0}_{\mathrm{D}}=\overline{h_{ij}\pi^{ij}}
=h_{ij}\pi^{ij}-\frac{\sqrt{h}}{\int_{\Sigma_t}\sqrt{h}}
\int_{\Sigma_t}h_{ij}\pi^{ij}\approx0,\quad
\chi^{i}_{\mathrm{D}}=\partial_{j}\left(h^{ \frac{1}{3}}h^{ij}\right)
\approx0.
\end{equation}
The BRST invariant gauge and ghost action for these gauge conditions can
be written in the form
\begin{equation} \label{EAdirac}
 S^{UD}_{gf+gh}=\int d^4x\left( -\bar\eta\bar\chi_{\mathrm{D}}^0
-\eta_i\chi_{\mathrm{D}}^i -\bar{c}^{*}s_b\bar\chi_{\mathrm{D}}^0
-c^{*}_i s_b\chi_{\mathrm{D}}^i \right),
\end{equation}
where the pair of ghosts $\bar{c},\bar{c}^{*}$ are average-free, while
the ghosts $c^i,c^{*}_j$ are not.

Let us start by computing the Slanov variation of the gauge conditions
$\chi^{i}_{\mathrm{D}}$. This demand some direct calculation that
results into
\begin{align}\label{s_b.uDiraci}
 s_b\chi^{i}_{\mathrm{D}} &=-2\kappa\partial_j\left[
h^{-\frac{1}{6}}\left(
 \pi^{ij} -\frac{1}{3}h^{ij}h_{kl}\pi^{kl} \right)\bar{c} \right]
+\frac{2}{3}\chi^{i}_{\mathrm{D}}D_jc^j   -\chi^{j}_{\mathrm{D}}D_jc^i
\nonumber \\
 & \quad -h^{\frac{1}{3}}\left( \delta^i_j h^{kl}\partial_k\partial_l
 +\frac{1}{3}h^{ik}\partial_k\partial_j \right)c^j.
\end{align}
Next we proceed to compute the Slanov variation of the gauge condition
$\bar\chi^{0}_{\mathrm{D}}$,
\begin{equation}
 s_b\bar\chi^{0}_{\mathrm{D}}=s_b\left(h_{ij}\pi^{ij}\right)
-s_b\sqrt{h}\left( \frac{1}{\int_{\Sigma_t}\sqrt{h}}
\int_{\Sigma_t}h_{ij}\pi^{ij} \right)  -\sqrt{h}s_b\left(
\frac{1}{\int_{\Sigma_t}\sqrt{h}} \int_{\Sigma_t}h_{ij}\pi^{ij} \right),
\label{s_b.uDirac0}
\end{equation}
where the last term of the above expression drops out of the action
\eqref{EAdirac}, since the ghost $\bar{c}^*$ has a vanishing average.

After evaluating the respective variation, we can use the resulting
expression\eqref{s_b.uDirac0} in order to write the third term of the
action \eqref{EAdirac} in the following form
\begin{align}\label{Sgh0.uDirac}
 \int d^4x\bar{c}^{*}s_b\bar\chi_{\mathrm{D}}^0  &=\int
d^4x\bar{c}^{*}\left( \frac{3}{2}\bcH_T\bar{c}
-\frac{2}{\kappa}\sqrt{h}\left( D^iD_i-\sR \right)\bar{c}
 +\partial_k\left( h_{ij}\pi^{ij}c^k \right) \right) \nonumber \\
 &\quad+3\int d^4x\sqrt{h}\bar{c}^{*}\bar{c}  \left(
\frac{1}{\int_{\Sigma_t}\sqrt{h}}
 \int_{\Sigma_t}\left[ \frac{\kappa}{\sqrt{h}}
\pi^{ij}\cG_{ijkl}\pi^{kl} -\frac{\sqrt{h}}{\kappa} \sR \right] \right)
\nonumber  \\
 &\quad-\int d^4x\bar{c}^{*} \left( -\frac{\kappa}{2}h_{ij}\pi^{ij}\bar{c}
+\sqrt{h}D_ic^i \right) \left[
\frac{1}{\int_{\Sigma_t}\sqrt{h}}\int_{\Sigma_t}h_{ij}\pi^{ij}  \right].
\end{align}
This is a problematic result, since it contains quadratic terms in
$\pi^{ij}$ that are not constraints. In the path integral, the
Faddeev-Popov determinant should be at most linear in $\pi^{ij}$ so that
the (gaussian) integration over the momenta can be performed. The quadratic terms
should involve a constraint so that they can be absorbed via shifts of
Lagrange multipliers. Above only the constraint term $\frac{3}{2}\bcH_T$
appears, while the integrated term is not a constraint. Indeed we could
use the constraint $\cH_0$ to write
\begin{equation}
 \frac{1}{\int_{\Sigma_t}\sqrt{h}}  \int_{\Sigma_t}\left(
\frac{\kappa}{\sqrt{h}}
 \pi^{ij}\cG_{ijkl}\pi^{kl} -\frac{\sqrt{h}}{\kappa} \sR \right)
 =\frac{\cH_0}{\int_{\Sigma_t}\sqrt{h}}-\lambda_0,
\end{equation}
but then the cosmological constant variable $\lambda_0$ reappears, which
is not correct since it is integrated in the path integral to obtain the
averaged unimodular condition factor $\delta\left(\int_{\Sigma_t}\left(
\sqrt{-g}-\epsilon_0 \right) \right)$ \cite{Bufalo:2015wda}. The last term in
\eqref{Sgh0.uDirac} is equally problematic, since it also involves a
quadratic $\pi^{ij}$ term, which is not a constraint.

\end{document}